\documentclass[twocolumn,showpacs,preprintnumbers,amsmath,amssymb]{revtex4}

\usepackage{graphicx}
\usepackage{dcolumn}
\usepackage{amssymb}
\usepackage{bm}
\usepackage{amsmath} 
\usepackage{color}
\usepackage{eurosym}

\begin{document}

\medskip

\title{Exact edge landing probability for the bouncing coin toss and the three-sided die problem}

\author{Llu\'is Hern\'andez-Navarro$^{1,4}\footnote{lluishn@gmail.com}$ and Jordi Pi\~nero$^{2,3,4}\footnote{jpinerfe@gmail.com}$}
\affiliation{$^{1}$Institut d'Investigacions Biom\`ediques August Pi i Sunyer (IDIBAPS), Barcelona, Spain}
\affiliation{$^{2}$ICREA-Complex Systems  Lab, Universitat Pompeu Fabra, 08003 Barcelona, Spain}
\affiliation{$^{3}$Institut de Biologia Evolutiva (CSIC-UPF), Psg Maritim Barceloneta, 37, 08003 Barcelona, Spain}
\affiliation{$^{4}$The Rashevsky Club, Barcelona, Spain}

\date{\today}

\begin{abstract}
Have you ever taken a disputed decision by tossing a coin and checking its landing side? This ancestral ``heads or tails'' practice is still widely used when facing undecided alternatives since it relies on the intuitive fairness of equiprobability. However, it critically disregards an interesting third outcome: the possibility of the coin coming at rest on its edge. Provided this evident yet elusive possibility, previous works have already focused on capturing all three landing probabilities of thick coins, but have only succeeded computationally as an exact and fully grounded analytical solution still evades the physics community. In this work we combine the classical equations of collisions with a statistical--mechanics approach to derive the exact analytical outcome probabilities of the partially inelastic coin toss. We validate the theoretical prediction by comparing it to previously reported simulations and experimental data; we discuss the moderate discrepancies arising at the highly inelastic regime; we describe the differences with previous, approximate models; and we finally propose the optimal geometry for the fair, cylindrical three-sided die based on the current results.
\end{abstract}


\maketitle

\section{Introduction}

Since long ago, and despite the apparent simplicity of the process involved, a great number of scientist have struggled to solve the problem of the face landing probabilities in real, physical tosses. Several of the first written documents where this question was confronted date back from the XVII century, when Galileo Galilei, Gerolamo Cardano and Christiaan Huygens wrote about the die throw and the coin toss problem~\cite{galilei1898sopra,cardano1953book}; and Blaise Pascal and Pierre Simon de Fermat exchanged letters discussing dice games ~\cite{de2007geometria}. Perhaps, the essence of the problem was best put forward by Newton in a private communication to Whiteside: ``{\em If a die bee not a regular body but a Parallelepipedon or otherwise unequally sided, it may bee found how much one cast is more easily gotten then another.}''~\cite{whiteside1967mathematical}.

Despite the popularity of the topic, reported attempts to solve this problem came more than a century later, by the hand of Simpson in 1740~\cite{simpson1740nature}. In this first model, only geometrical arguments were taken into account when computing the outcome probabilities, which were assumed proportional to the solid angle of each face. As such, this model only approximately captured the outcome of a perfectly inelastic toss with no angular momentum.

Simpson's model has been extensively rejected for non-inelastic tosses, as reported in the experimental works of Buden~\cite{budden198064}, Singmaster~\cite{singmaster198165} and Heilbronner~\cite{heilbronner1985crooked}, but it is still adopted in current scientific outreach~\cite{parker2018thick} and in recent studies due to its simplicity~\cite{bosso2021experimental}. Nevertheless, this model is more often presented in an extended version where not only the geometry but also the angular momentum are taken into account~\cite{keller1986probability,diaconis2007dynamical,yong2011probability,aciun2020modeling}, providing a solution for the non--bouncing toss only.

Beyond Simpson's geometrical approach and its extensions to solve the non-bouncing problem, two new classes of models have arisen in the last few decades to assess more realistic, bouncy tosses. On one hand, probabilistic models grounded on the chaotic properties and randomness of the coin toss~\cite{keller1986probability,strzalko2008dynamics} have been suggested; as well as heuristic models are based on the Gibbs distribution from equilibrium thermodynamics~\cite{riemer2014cuboidal}, or on Markovian processes with heuristic transition rates~\cite{kuindersma2007teaching,pender2014predicting}. On the other hand, dynamical models strictly grounded on the equations of motion for partially inelastic collisions~\cite{murray1993probability,bondi1993dropping,kapitaniak2012three} have also been proposed.

As for the dynamical approaches, a computational model has proven successful to capture even partially inelastic, bouncing tosses of short cylinders, i.e. coins, where the heads, tails and edge landing probabilities were assessed~\cite{murray1993probability}. Yet, provided the cumulative complexity with each bounce, an exact analytical solution for the toss of a cylinder has only been found at most for up to two bounces, as reported in the work of~\cite{bondi1993dropping}.

Here we present the exact solution of the partially inelastic coin toss, grounded on the equations of collisions from classical mechanics~\cite{murray1993probability}, and tackled by means of non-equilibrium statistical mechanics under {\it a priori} equiprobability of phase states. The latter is simply achieved by highly random initial linear and angular velocities in the toss, and/or by the randomization of these variables after enough partially inelastic collisions take place. The final analytical expression is provided and validated from reanalyzed experimental and numerical data; and the specific case of the fair three-sided cylindrical die is discussed.

\section{Exact Model}

In order to tackle the complex coin toss problem analytically, some simplifications have to be adopted. In this section we will first describe the initial assumptions necessary to solve our problem, while ensuring that the essential features of the real-world problem are not lost or oversimplified. We will also observe the key variables that fully describe the instantaneous phase state of the coin under those assumptions. Grounded on all of the above, we will provide the analytical step-by-step derivation of the landing probabilities for the coin toss.

\subsection{Initial assumptions and approach}
\label{assumptions}

\begin{figure}
\includegraphics[width=\columnwidth]{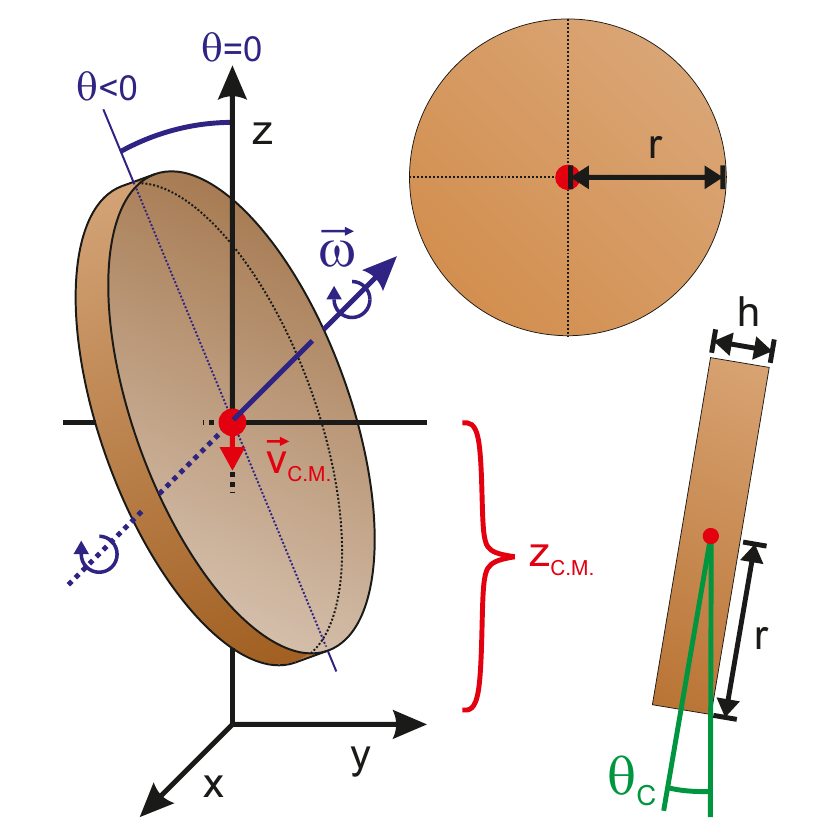}
\caption{{\em Coin sketch.} The coin, a homogeneous and incompressible cylinder of radius \(r\) and height \(h\), is tossed with an initial CM height \(z\cdot\widehat{a_z}\), and with vertical velocity \(v\cdot\widehat{a_z}\). The coin's initial orientation with respect to the z-axis is depicted by the polar angle \(\theta\cdot\widehat{a_\theta}\), with related angular velocity \(-\omega\cdot\widehat{a_x}\). Since air and ground frictions are neglected, and the toss starts with no angular velocity for the two remaining principal axes, the dynamics are fully captured by \(\{z, v; \theta, \omega\}\) alone. The {\em Critical angle} \(\theta_c\) corresponds to the maximum polar angle at which the coin doesn't tip over its edge.}
\label{sketch}
\end{figure}

The coin is conceptualized as a perfect, incompressible cylinder of radius $r$ and height $h$ (Fig.~\ref{sketch}). The tossing imbues a linear velocity $\vec{v}$ and an angular speed $\vec{\omega}$ to the coin such that the first vector is purely vertical (i.e. z-axis component only), and the latter vector is applied to a rotational axis that crosses the coin's Center of Mass (CM) on it's diameter parallel to the floor (i.e. x-component only). The time-dependent angle of the coin around the x-axis is depicted by $\theta$. Additionally, the coin bounces multiple times on the floor in successive inelastic collisions (characterized by a constant coefficient of restitution $\gamma$) before reaching its equilibrium state, and with a given dissipation of energy at each bounce. Further dissipative forces such as air and ground frictions are neglected. Due to the chaotic-like properties of the coin toss, states in the phase space are assumed to be independent of the initial conditions as long as enough bounces take place~\cite{strzalko2008dynamics,kapitaniak2012three}. Therefore, as long as the previous condition holds, we may consider equiprobability of states in the phase space.

\subsection{Model derivation}

The unbiased coin toss has long been regarded as the paradigm of a random binary variable, with an exactly equal probability for either heads or tails. Although these two final stable states are indeed equally probable (in the ideal scenario), the previous conceptualization of the coin toss is not strictly true: real coins present a third, less frequent {\em meta}stable state corresponding to resting on its edge.

To fully characterize the outcome probability distribution of a toss, the probability of a coin landing on its edge has to be properly addressed. As already discussed in the introduction, inspiring effort has been previously invested on this topic, as it is for instance shown in the computational work of~\cite{murray1993probability}. On the other hand, very few analytical steps, involving full statistical physics-driven solutions for the problem, have been put forward before~\cite{bondi1993dropping}. The current paper offers a novel systematic and purely analytical approach to tackle the coin toss problem and potentially other non-equilibrium dissipative systems.

\subsubsection{Potential, kinetic and minimum potential energy}

\begin{figure}
\includegraphics[width=\columnwidth]{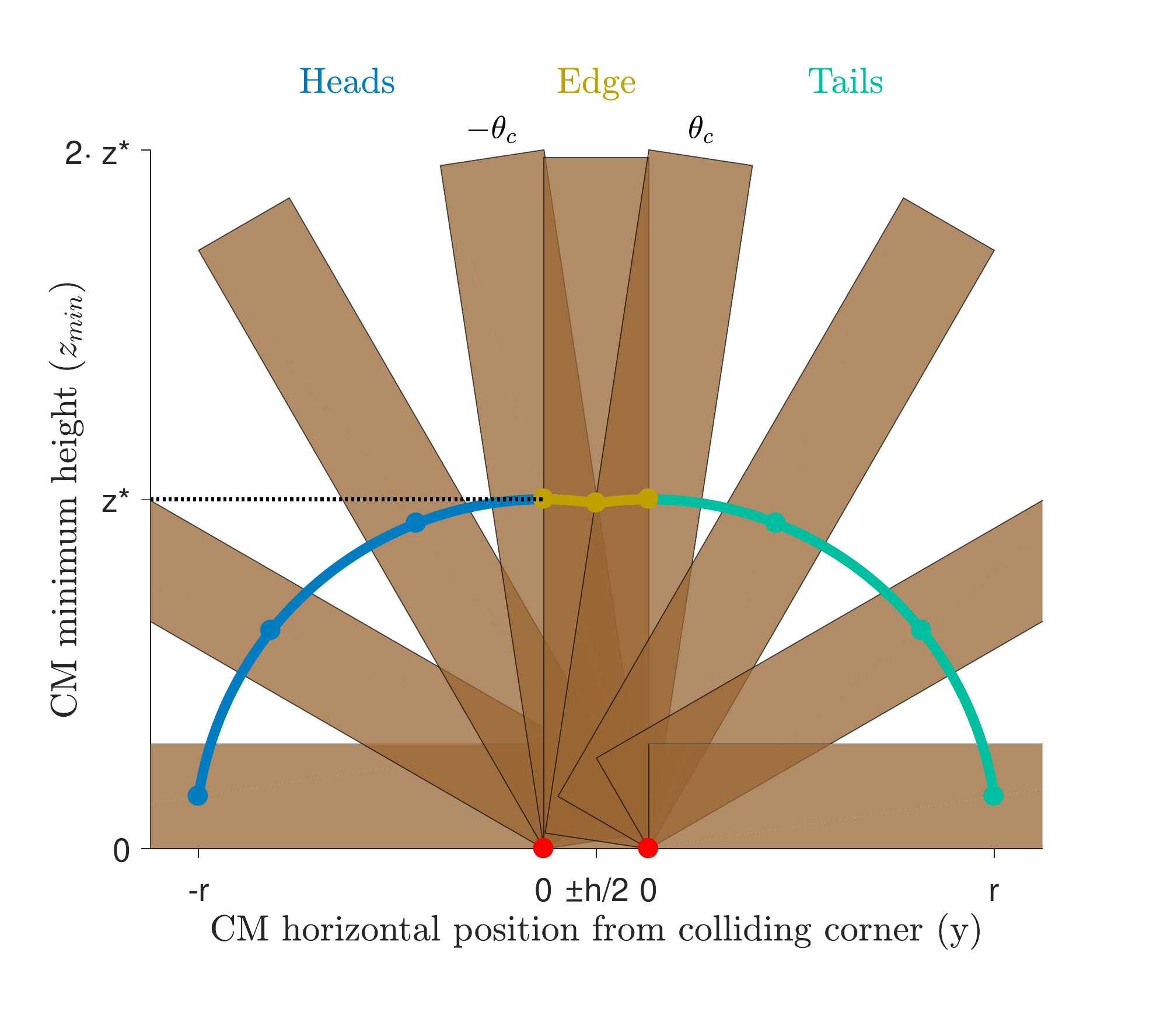}
\caption{{\em Minimum potential energy.} The minimum potential is proportional to the height of the CM when at least one of the corners of the coin is in contact with the ground, and is set by the angle \(\theta\), as depicted by the solid line. Solid circles on the line depict the position of the CM for each instance of the polar angle. Different colors show the instantaneous outcome of the toss, set by the domain of the current \(\theta\). When the mechanical energy is lower than the critical value (\(E<E_c=mgz^*\), Eq.~\ref{E_c}), the coin cannot overcome the potential energy at all angles, the angle domains become disjoint, and the final outcome of the toss is bound to its current instantaneous value. Red circles depict the coin--ground contact points, i.e. the colliding corners.}
\label{minimum_potential}
\end{figure}

The potential energy $U$ is directly determined by the vertical height $z$ of the CM of the coin as $U=mgz$, where $m$ is the mass of the coin, and $g$ is the acceleration of gravity. The kinetic energy corresponds to
\begin{eqnarray}
T&=&\frac{m v^2}{2}+\frac{I_{xx} \omega^2}{2} \ ,
\label{T}
\end{eqnarray}
where $I_{xx}$ is the moment of inertia of the coin around the rotational axis described previously. Recall here our initial assumption on the angular momentum being constrained to the x--axis only.
To address the final state of the coin after the toss, it will be useful to define the minimum potential energy $U_{min}(\theta)=m gz_{min}(\theta)$ as the profile of the potential energy depending on $\theta$ when at least one corner of the coin is in contact with the floor (Fig.~\ref{minimum_potential}). Under the previous constraint, and assuming coin's homogeneity, the height of the CM reads
\begin{eqnarray}
z_{min}\left(\theta\right)&=&\sqrt{r^2+\left(h/2\right)^2}  \cos\left(\left|\theta\right|-\arctan\left(\frac{h}{2r}\right)\right) \nonumber \\
&=&z^* \cos\left(\left|\theta\right|-\theta_c\right),
\label{z_min}
\end{eqnarray}
where $z^*$ is the semi-diagonal of the cross section of the coin (i.e. maximum CM height possible when the coin is in contact with the floor), and $\theta_c\equiv \arctan\left(\frac{h}{2r}\right)$ is the critical angle of the coin (see Figs.~\ref{sketch} and~\ref{minimum_potential}).

\subsubsection{Critical energy}

In a coin toss, non-adiabatic dissipative forces are at play during each partially inelastic collision with the floor. Therefore, the system loses energy discontinuously over time in a staircase-like fashion, up to the final (meta)stable state. We define the critical energy $E_c$ as the energy threshold below which the final outcome of the toss is fully determined by the local domain of the instantaneous angle $\theta\left( t\right)$ alone (Fig.~\ref{minimum_potential}). That is, at lower mechanical energy than $E_c$, the coin is not able to overcome the potential energy barrier that separates its now disjoint \(\theta\) domains. Thus, \(\theta\left( t\right)\) is constrained to its current local domain, and will gradually be dragged to the local minimum energy state (either heads, tails or edge) as the coin loses energy with each successive collision.

The critical energy $E_c$ is directly set by the maximum potential energy of the coin when it is in contact with the floor, which corresponds to
\begin{eqnarray}
E_c&=&m g z^*,
\label{E_c}
\end{eqnarray}
where $z^*$ is again the maximum possible height of the CM of the coin when in contact with the floor.

\subsubsection{Critical collision}
\label{critical_collision}

We define the $critical~collision$ as the single collision at which the energy of the coin becomes lower than $E_c$. Since the angle $\theta$ does not change during the instantaneous collision, its value in the immediately prior state to the critical collision already determines the final outcome of the toss. Being so, the edge landing probability can be derived as the fraction of prior states in the phase space with $\theta$ in the edge domain (i.e. $-\theta_c\leq\theta\leq\theta_c$).

The states immediately prior to the critical collision are constrained by a physical contact between the coin and the floor (where $z_{CM}$ is determined by $\theta$; see Eq.~(\ref{z_min})). Moreover, the distribution of states is governed by the equations of the partially inelastic collision, derived on the following section.

\subsubsection{Partially inelastic collision}

The partially inelastic collisions are fully characterized by the restitution coefficient $\gamma$ and the equation
\begin{eqnarray}
u''&=&-\gamma u',
\label{inelastic_collision}
\end{eqnarray}
where $u'$ and $u''$ are the vertical velocity of the colliding corner of the coin before and after the collision, respectively.

The corner's vertical velocity change $\Delta u$ is derived by adding $-u'$ to both sides of the previous Eq.~(\ref{inelastic_collision}) and rearranging terms:
\begin{eqnarray}
\Delta u&=&-(1+\gamma) u'.
\label{change_u}
\end{eqnarray}

On the other hand, changes in the corner's vertical velocity can be directly related to changes in both the CM's vertical velocity $\Delta v$ and in the angular momentum of the coin $\Delta\omega$ as
\begin{eqnarray}
\Delta u&=&\Delta v+y(\theta)\Delta\omega,
\label{change_v_w}
\end{eqnarray}
where $y(\theta)$ depicts the horizontal position of the coin's CM with respect to the colliding corner (see Fig.~\ref{minimum_potential}; note that we adopted the opposite sign convention for \(\theta\), \(y\) and \(\omega\) than in \cite{murray1993probability}).

Additionally, we apply conservation of angular momentum using a reference axis that crosses the contact point between coin and floor (i.e. the colliding corner), and that is parallel to the x-axis of the problem. Conservation of angular momentum only holds for this reference axis since it is the only axis where the torque applied by the floor to the coin vanishes. The angular momentum is then the rotational movement around the CM, as well as the translational movement of the coin's CM projected as a rotation around the colliding corner, which yields
\begin{eqnarray}
m(-y)\Delta v+I\Delta\omega&=&0,
\label{conservation_L}
\end{eqnarray}
where $I\equiv I_{xx}$ to simplify the notation.

By combining Eqs.~(\ref{change_u}, \ref{change_v_w}, and \ref{conservation_L}) we obtain the final expressions for the changes in the velocity of the CM and in the angular velocity
\begin{eqnarray}
\Delta v&=&-(1+\gamma)\cdot \frac{I}{I+m y^2}\left(v'+y\omega'\right),
\label{change_v}
\end{eqnarray}
\begin{eqnarray}
\Delta\omega&=&-(1+\gamma) \cdot  \frac{m y}{I+m y^2}\left(v'+y\omega'\right).
\label{change_w}
\end{eqnarray}
\\
And finally, combining Eq.~(\ref{T}) with Eqs.~(\ref{change_v} and \ref{change_w}), we derive the energy loss of the coin at a single bounce:
\begin{eqnarray}
\Delta E&=&-\frac{1-\gamma^2}{2}\cdot \frac{mI}{I+my^2}\left(v'+y\omega'\right)^2.
\label{change_E}
\end{eqnarray}

Note that all results are consistent with the reduced equations used in the computational study of~\cite{murray1993probability}.

\subsubsection{Phase space density of prior states}

By definition, states immediately prior to the critical collision (where $\theta$ already determines the final outcome of the toss) fulfill three conditions. First, their energy must be greater than or equal to $E_c$; second, their corresponding energy immediately after the collision has to be lower than $E_c$, following Eq.~(\ref{change_E}); and third, \(u'\) has to be negative by physical constraints. The previous conditions are summarized by the following inequalities:
\begin{eqnarray}
E'\left(\theta,v,\omega\right)&\geq&E_c,
\label{E_constraint_1.0}
\end{eqnarray}
\begin{eqnarray}
E_c&>&E''\left(\theta,v,\omega\right)=E'\left(\theta,v,\omega\right)+\Delta E,
\label{E_constraint_2.0}
\end{eqnarray}
\begin{eqnarray}
\text{and }~v'+y\cdot\omega'<0.
\label{E_constraint_3.0}
\end{eqnarray}

The explicit version of the former Eq.~(\ref{E_constraint_1.0}) becomes
\begin{eqnarray}
m g z_{min}(\theta)+\frac{m v^2}{2}+\frac{I\omega^2}{2}&\geq&m g z^*,
\label{E_constraint_1.1}
\end{eqnarray}
which can be rearranged to an inequality whose active constraint (case `=') follows an elliptic surface
\begin{eqnarray}
&\left[\frac{1}{2g\left(z^*-z_{min}\right)}\right] v^2+\left[\frac{I}{2mg\left(z^*-z_{min}\right)}\right]\omega^2\geq1.
\label{E_constraint_1.2}
\end{eqnarray}

On the other hand, Eq.~(\ref{E_constraint_2.0}) becomes
\begin{eqnarray}
&mg z_{min}(\theta)+\frac{m v^2}{2}+\frac{I\omega^2}{2}+  \nonumber \\
&-\frac{\left(1-\gamma^2\right) mI}{2\left(I+my^2(\theta)\right)}\cdot\left(v+y(\theta)\omega\right)^2<m g z^*,
\label{E_constraint_2.1}
\end{eqnarray}
which can be rearranged into an inequality with a constraint surface that follows a tilted ellipse
\begin{eqnarray}
&\left[\frac{I\gamma^2+my^2}{2g\left(I+m y^2\right)\left(z^*-z_{min}\right)}\right]\cdot v^2
+\left[\frac{Iy^2\gamma^2+I^2/m}{2g\left(I+m y^2\right)\left(z^*-z_{min}\right)}\right]\cdot\omega^2 \nonumber \\
&+\left[\frac{Iy\left(1-\gamma^2\right)}{-g\left(I+m y^2\right)\left(z^*-z_{min}\right)}\right]\cdot v\omega
<1.
\label{E_constraint_2.2}
\end{eqnarray}

\begin{figure}
\includegraphics[width=\columnwidth]{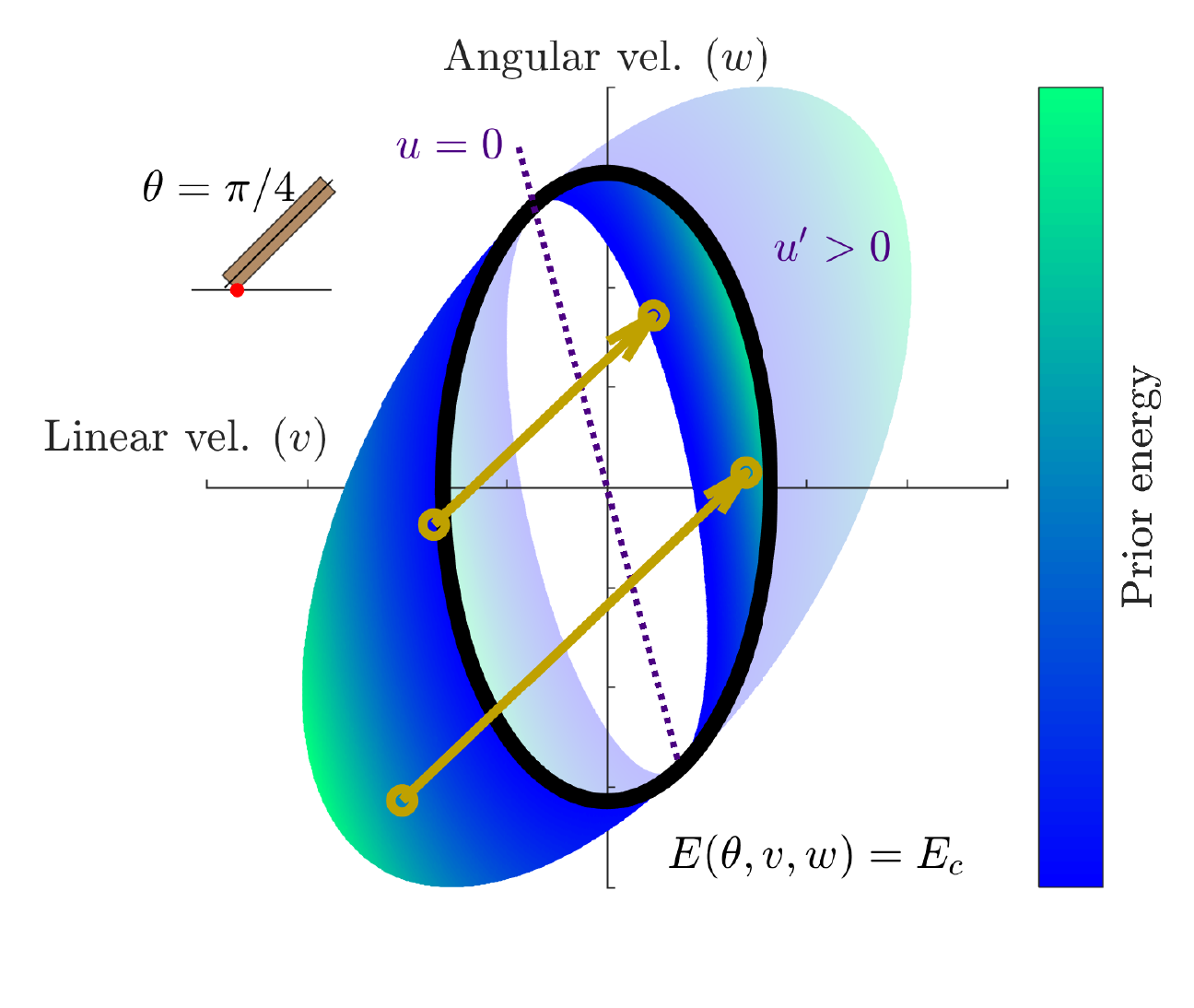}
\caption{{\em Phase space prior and posterior to critical collisions.} Phase space in \(v\) and \(\omega\) coordinates for an example polar angle (\(\theta=\pi/4\)), coin geometry (\(\theta_c=\pi/6\)) and coefficient of restitution (\(\gamma=0.5\)) during a (critical) collision. Black circumference illustrates the critical energy boundary of Eq.~(\ref{E_constraint_1.2}). Dashed purple line shows the limiting case \(u=0\). Filled solid area encircled between the outer tilted ellipse of Eq.~(\ref{E_constraint_2.2}) and \(E_c\) depicts all the {\em prior states} that fulfill \(u'<0\) and will become subcritical after a critical collision; transparent area illustrates non-physical \(u'>0\). Filled solid surface comprised between \(E_c\) and the inner tilted ellipse encloses all the phase states that are possible immediately after a critical collision; transparent surface depicts non-physical \(u''<0\). The inner tilted ellipse is obtained by reversing time with \(\gamma\rightarrow\gamma^{-1}\) in Eq.~(\ref{E_constraint_2.2}). The angle of tilt for both prior and posterior ellipses are set by the collision angle \(\theta\). Golden arrows depict two example critical collisions, with the corresponding transitions from prior to posterior phase states. Colormap illustrates the kinetic energy of the prior states, for both the prior states itself and the resulting posterior states. Inset: collision sketch.}
\label{phase_space}
\end{figure}

Figure~\ref{phase_space} shows an illustrative sketch of the phase subspace compatible with prior states for an example $\theta$, \(\theta_c\), and \(\gamma\) on a critical collision.

Since, for prior states, $\theta$ alone fully determines the outcome of the toss (and also the CM's height $z_{min}$), we derive the density function of prior states depending only on this variable. To do so, we integrate the prior states' phase subspace over all possible values of the remaining parameters, which are $p=m v$ and $L=I\omega$, the linear and angular momentum just before the critical collision, respectively. We note that, because we are directly integrating the constrained 2--D volume of this subspace, we are implicitly assuming equiprobability of prior states, as discussed in section~\ref{assumptions}.

As shown for an example case in Fig.~\ref{phase_space}, the aforementioned integration is equivalent (up to a phase space normalization factor) to computing the area delimited by the tilted ellipse of Eq.~(\ref{E_constraint_2.2}), subtracting the area delimited by the ellipse of the complementary of Eq.~(\ref{E_constraint_1.2}), and halving the result to capture \(u'<0\). Hence, we take into account all prior states at fixed \(\theta\) that will have less energy than \(E_c\) after the collision, and discard those that already have below-critical energy.

The general formula for the area of an ellipse of the form $A\cdot X^2+B\cdot Y^2+C\cdot X Y\leq1$ is
\begin{eqnarray}
S=\frac{2\pi}{\sqrt{4A B-C^2}}.
\label{area_ellipse}
\end{eqnarray}

Then, the inner area delimited by the complementary of Eq.~(\ref{E_constraint_1.2}) is simply
\begin{eqnarray}
S_1=2\pi g\sqrt{\frac{m}{I}}\cdot\left(z^*-z_{min}(\theta)\right).
\label{S_1}
\end{eqnarray}

As for the second constraint of prior states, the area delimited by Eq.~(\ref{E_constraint_2.2}) is
\begin{eqnarray}
S_2=\frac{2\pi g\left(z^*-z_{min}(\theta)\right)\left(I+my^2\right)}
{\sqrt{\left(I\gamma^2+my^2\right)\left(Iy^2\gamma^2+\frac{I^2}{m}\right)-I^2y^2\left(1-\gamma^2\right)^2}}~~
\label{S_2.0}
\end{eqnarray}
which becomes
\begin{eqnarray}
S_2=\frac{2\pi g\left(z^*-z_{min}(\theta)\right)\left(I+m y^2\right)}
{\sqrt{\left(I\gamma^2/m\right)\left(I^2+m^2y^4+2Imy^2\right)}},
\label{S_2.1}
\end{eqnarray}
which reduces to
\begin{eqnarray}
S_2=2\pi g\sqrt{\frac{m}{I}}\cdot\frac{\left(z^*-z_{min}(\theta)\right)}{\gamma},
\label{S_2.2}
\end{eqnarray}
where the variable $y(\theta)$ is canceled out, and the angle only contributes in $z_{min}$ as in Eq.~(\ref{S_1}).

Finally, the density function of the prior states $\rho$ in terms of $\theta$ is
\begin{eqnarray}
\rho(\theta)&=&\frac{m I}{h^2}\frac{\left(S_2-S_1\right)}{2} \nonumber \\
&=&\frac{\pi m^{3/2} g \sqrt{I}}{h^2}\cdot\frac{1-\gamma}{\gamma}\cdot\left(z^*-z_{min}(\theta)\right),~
\label{S_2.3}
\end{eqnarray}
where $h$ is the Planck constant. The multiplicative prefactor $m\cdot I/h^2$ is introduced in order to properly normalize the phase space (composed of two pairs of conjugated variables: \{$z$, $p=mv$\} and \{$\theta$, $L=I\omega$\}). However, note that this factor cancels out during the computation of the outcome probabilities in the next section.

\subsubsection{Edge landing probability}

As previously discussed in section~\ref{critical_collision}, the probability of a coin toss landing on edge can be derived as the fraction of prior states in the phase space whose angle $\theta$ fulfills $-\theta_c\leq\theta\leq\theta_c$ (see Fig.~\ref{minimum_potential}). Due to the coin's symmetry, and provided that here we are not interested on distinguishing `heads' from `tails' in non-edge outcomes, we can collapse the whole $\theta$ domain to the first quadrant. Therefore, the edge landing probability $P_E$ yields
\begin{eqnarray}
P_E&=&\frac{\int_{\theta=0}^{\theta=\theta_c}\rho(\theta) d\theta}{\int_{\theta=0}^{\theta=\pi/2}\rho(\theta) d\theta}
=\frac{\int_{0}^{\theta_c}\left(1-\cos(\theta-\theta_c)\right) d\theta}{\int_{0}^{\pi/2}\left(1-\cos(\theta-\theta_c)\right) d\theta} \nonumber \\
&=&\frac{\theta_c-\sin(\theta_c)}{\pi/2-\left(\sin(\theta_c)+\cos(\theta_c)\right)},
\label{P_E}
\end{eqnarray}
that, interestingly, only depends on the geometry of the coin (i.e. $\theta_c$), and not on the restitution coefficient $\gamma$. For the examples of 1 \pounds, 1 \euro, and a quarter \$ coins, the theory predicts an edge outcome probability of 1 over \(\sim\)1000, 3000, and 8000 tosses, respectively.

A Taylor expansion of the above result for small $\theta_c$ provides the asymptotic edge landing probability
\begin{eqnarray}
P_E\sim\frac{{\theta_c}^3}{3\left(\pi-2\right)}.
\label{P_E_Taylor}
\end{eqnarray}

\section{Discussion}

\begin{figure}
\includegraphics[width=\columnwidth]{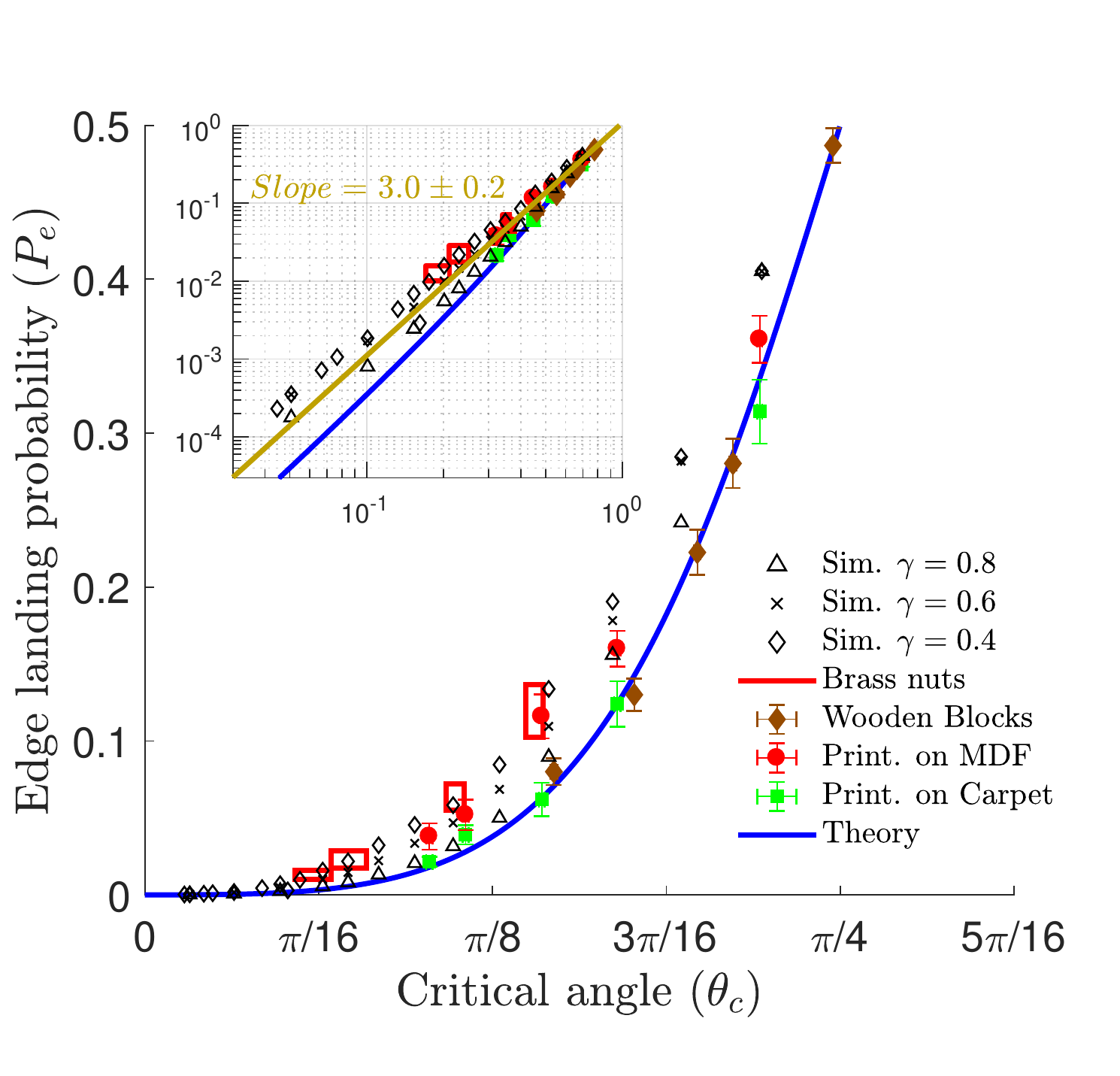}
\caption{{\em Theoretical, numerical and experimental results.} Comparison of the theoretical prediction to the experimental and simulated data of two preexisting datasets. Solid symbols indicate the experimental data reported in~\cite{pender2014predicting} for the toss of wooden blocks and 3D--printed blocks on tough carpet (brown diamonds and red circles, respectively), and printed blocks on Medium Density Fibreboard (MDF, green squares). Error bars show the binomial error. Open red rectangles depict the experimental data of~\cite{murray1993probability} for the toss of brass nuts, where their width and height show one standard deviation around the mean; open symbols indicate their numerical simulations performed for distinct values of the coefficient of restitution \(\gamma\), with error bars smaller than symbols. Solid blue line depicts the theoretical prediction of Eq.~(\ref{P_E}), which only depends on the geometry of the coin, and through the critical angle \(\theta_c\). Inset: data in logarithmic scale; the golden line depicts the power-law fit to all data, with an exponent of \(3.0\pm0.2\), consistent with the prediction of Eq.~(\ref{P_E_Taylor}) for small \(\theta_c\).}
\label{results}
\end{figure}

To test the exact analytical solution for the partially inelastic coin toss derived in this manuscript, we compare the prediction of Eq.~(\ref{P_E}) to two different preexisting datasets, and present the results in Fig.~\ref{results}. First, since the initial approach and assumptions of this study are grounded on and shared by the work of~\cite{murray1993probability}, we compare our exact solution to their numerical simulations of the coin toss, as well as to their experimental data of brass nuts tosses. And second, we take into account the experimental data of~\cite{pender2014predicting}, where the toss of very long cuboids is studied. Note that, although the geometrical shapes of a coin and a long cuboid are radically different, the dynamics of their respective tosses are very similar. This is because, during the toss, the long cuboid almost always rotates around its longest axis only, effectively behaving as a coin (or as a two-dimensional die, as discussed in~\cite{pender2014predicting}).

On one hand, as shown in Fig.~\ref{results}, the exact solution neatly captures the edge landing probability for the toss of wooden blocks (cuboids) and 3D printed blocks of different sizes and proportions falling on tough carpet. On the other hand, the theoretical prediction moderately underestimates the probability of edge outcomes for the toss of brass nuts and of printed blocks falling on a Medium Density Fibreboard (MDF) surface, where they tend to `clatter' in a more inelastic regime. Lastly, numerical simulations of the coin toss show that the more elastic the toss is, the better the exact solution captures the observed edge outcome probability.

The exact model predicts no dependency of the outcome probabilities on the coefficient of restitution \(\gamma\), i.e. it predicts the same outcome probabilities for different inelastic regimes as long as the geometry of the coin is conserved. However, both experiments and simulations show a moderate dependency of the edge landing probability with the degree of inelasticity of the toss. As discussed in~\cite{murray1993probability}, the good agreement between the simulations and experimental data validates the initial assumptions for the real-case scenario, shared by both simulations and theory. Therefore, a constant coefficient of restitution \(\gamma\), as well as an incompressible coin, negligible air and ground friction, and angular momentum limited to the z--axis, are valid assumptions for real tosses. In fact, any discrepancy between the experimental results in the highly inelastic limit and our theoretical prediction can only arise from the assumption of equally probable prior states.

The theoretical assumption of equiprobability in the phase space is grounded on the chaotic properties of the coin toss, which have been shown to hold better the higher the number of collisions before the coin sets~\cite{vulovic1986randomness,strzalko2008dynamics}. The limited initial energy of the simulated tosses to avoid ``increases in computing time''~\cite{murray1993probability}, together with a limited randomization of the initial conditions (zero initial linear and angular velocities), could hinder the chaotic nature of the process in the simulations, specially for the case of highly inelastic collisions where the number of bounces is greatly reduced ~\cite{kapitaniak2012three}. This limitation in the number of bounces is also shared by the highly inelastic tosses of printed blocks on MDF, but not by the printed and wooden blocks on tough carpet, ``a surface on which they roll, often through many revolutions, rather than `clattering'.''~\cite{pender2014predicting} (Fig.~\ref{results}, solid symbols with error bars). In conclusion, the theoretical prediction captures the outcome probability of any partially inelastic coin toss provided that the coin bounces enough times before coming to rest, a condition more easily fulfilled in real settings the more elastic the tosses are. 

Regarding alternative models, on one hand, it has been shown that experimental data does not reject the {\it a priori} assumption of adopting the Gibbs distribution for the outcome states in the probabilistic approach of~\cite{riemer2014cuboidal}. However, in addition to the previous heuristic assumption, the corresponding solution relies on a non-interpretable parameter \(\beta\), related to some kind of non-equilibrium statistical temperature. This \(\beta\) seems to depend (in an unknown way) on the material the tossed object and the colliding surface are made of, the friction between them and with the air, the inelasticity of the collisions, etc.; and the parameter can only be tied with the experimental data.

On the other hand, the probabilistic approach of the ``simple bounce'' model of~\cite{kuindersma2007teaching} relies on the assumption that a constant fraction of the kinetic energy is lost at each partially inelastic bounce, irrespective of the linear or angular momentum involved in the collision. Although this assumption greatly simplifies the problem, it is contrary to the standard approach in collision theory, where a constant restitution coefficient in the linear velocity of the colliding corner is usually considered~\cite{murray1993probability}. As expected, the model comparison conducted in their study rules out the ``constant relative energy loss'' assumption in favor of more flexible dependencies with dynamical parameters, consistent with our model. However, a deeper exploration within that study was hindered due to computational limitations.

Finally, regarding alternative models, it is important to acknowledge the contribution of~\cite{pender2014predicting}, who heuristically suggested that the transition probabilities from prior (supercritical) states to posterior (subcritical) states in the perfectly elastic limit might be proportional to the ``activation energy'' of each final outcome,  i.e. the difference between the critical energy and the lowest potential energy possible for the current \(\theta\) domain: heads, tails or edge. As we have proven in the full derivation of the exact solution, the above transition probabilities are in fact proportional to the difference between the critical energy and the angle-specific minimum potential energy \(mgz_{min}\left(\theta\right)\) for any partially inelastic collision, i.e. for any coefficient of restitution \(\gamma\). Although the perfectly elastic solution proposed by~\cite{pender2014predicting} was thus inaccurate in the strict conceptual sense, their analytical expression did quantitatively approximate the exact general solution of Eq.~(\ref{P_E}), and their insightful approach did point towards the right direction.

As for additional applications of our exact coin tossing solution, a natural follow--up is the derivation of the optimal geometry for the fair, cylindrical three-sided die. Indeed, one can simply constrain the edge landing probability in Eq.~(\ref{P_E}) to the fair value of one third, and then solve numerically for the critical angle that predicts this outcome probability (which provides the same 1/3 value for heads and tails too). Thus, the predicted critical angle for the fair three-sided die derived from our model is \(\theta_c^{fair}\simeq 0.693~\text{rad}\), hence, the ratio of cylinder height vs. diameter \(\eta:=h/2r \simeq 0.831\). We note that this value differs from previous approximations such as in~\cite{yong2011probability,parker2018thick,bosso2021experimental}, in which it was argued that \(\eta\simeq1/\sqrt{3}\simeq 0.577\). Just as in Simpson's model~\cite{simpson1740nature}, these approximations disregard the impact of bouncing occurring in real tosses, which is known to largely reduce the probability of edge outcome~\cite{murray1993probability}. On the other hand, the experimental data by Pender \& Uhrin (2014)~\cite{pender2014predicting} show a solid agreement with our theoretical prediction at the range of values for \(\theta_c\) that correspond to \(P_E\sim1/3\) (see Fig.~\ref{results}). We note that, strictly, our prediction for the three-sided die only holds for an initial angular momentum bounded around the x--axis (see Fig.~\ref{sketch}), or for no initial angular velocity (since collisions on a dropping cylinder imbue a rotation around the x--axis only~\cite{bondi1993dropping}). Hence, large initial angular momentum in any of the remaining principal axes of the three-sided die may introduce deviations from our theoretical prediction, effectively lowering \(\eta\) for the fair die.

In summary, in this study we have first derived the equations of motion and collision for the partially inelastic coin toss grounded on standard assumptions of classical mechanics. Second, we have used the previous equations to obtain constraints on the phase states prior and posterior to the critical collision, after which the outcome of the coin is set. Third, we have implemented a statistical--mechanics approach by computing analytically the number of prior phase states depending on each of the three possible outcomes: heads, tails and edge. And fourth, we have capitalized on the previous result to derive the exact analytical outcome probabilities of the partially inelastic coin toss. In the last section of the manuscript, we have validated the theoretical prediction by comparing it to previous simulations and experimental data; we have discussed the moderate discrepancies arising at the highly inelastic regime due to the limited number of bounces; we have highlighted the differences with previous, approximate models; and we have proposed the optimal geometry for the fair, cylindrical three-sided die at \(h/2r \simeq 0.831\).

To conclude, we believe that this work offers a detailed step-by-step derivation of the long sough exact solution of the coin toss problem, and opens the door to novel analytical approaches for tackling additional phenomena in non-equilibrium systems. Finally, we suggest that future extensions of this work might provide deep insight on the landing probabilities for biased coins, as well as for the general case of the three-dimensional die throw.

\begin{acknowledgments}
\small{The authors would like to thank all members of the {\em Rashevsky Club} for interesting and fruitful discussions, as well as acknowledge the inspiring work of Daniel Bruce Murray in the study of coin tossing. The authors declare that no funding was received for this project.}
\end{acknowledgments}
\vspace{0.5 cm}
\section*{Author Contributions}

L.H.-N. and J.P. conceived the original idea. L.H.-N. developed the analytical model, L.H.-N. and J.P. interpreted the results. L.H.-N. and J.P. wrote the manuscript.


\bibliographystyle{apsrev4-1}
\bibliography{bibliography.bib}

\end{document}